\PassOptionsToPackage{unicode}{hyperref}
\PassOptionsToPackage{hyphens}{url}
\documentclass[
  12pt,
  letterpaper]{article}
\usepackage{lmodern}
\usepackage{amssymb,amsmath}
\usepackage{ifxetex,ifluatex}
\ifnum 0\ifxetex 1\fi\ifluatex 1\fi=0 % if pdftex
  \usepackage[T1]{fontenc}
  \usepackage[utf8]{inputenc}
  \usepackage{textcomp} % provide euro and other symbols
\else % if luatex or xetex
  \usepackage{unicode-math}
  \defaultfontfeatures{Scale=MatchLowercase}
  \defaultfontfeatures[\rmfamily]{Ligatures=TeX,Scale=1}
\fi
\IfFileExists{upquote.sty}{\usepackage{upquote}}{}
\IfFileExists{microtype.sty}{% use microtype if available
  \usepackage[]{microtype}
  \UseMicrotypeSet[protrusion]{basicmath} % disable protrusion for tt fonts
}{}
\makeatletter
\@ifundefined{KOMAClassName}{% if non-KOMA class
  \IfFileExists{parskip.sty}{%
    \usepackage{parskip}
  }{% else
    \setlength{\parindent}{0pt}
    \setlength{\parskip}{6pt plus 2pt minus 1pt}}
}{% if KOMA class
  \KOMAoptions{parskip=half}}
\makeatother
\usepackage{xcolor}
\IfFileExists{xurl.sty}{\usepackage{xurl}}{} % add URL line breaks if available
\IfFileExists{bookmark.sty}{\usepackage{bookmark}}{\usepackage{hyperref}}
\hypersetup{
  pdftitle={The Future of Work Is Here},
  hidelinks,
  pdfcreator={LaTeX via pandoc}}
\usepackage{graphicx,grffile}
\makeatletter
\def\maxwidth{\ifdim\Gin@nat@width>\linewidth\linewidth\else\Gin@nat@width\fi}
\def\maxheight{\ifdim\Gin@nat@height>\textheight\textheight\else\Gin@nat@height\fi}
\makeatother
\setkeys{Gin}{width=\maxwidth,height=\maxheight,keepaspectratio}
\makeatletter
\def\fps@figure{htbp}
\makeatother
\title{The Future of Work Is Here}
\usepackage{etoolbox}
\makeatletter
\providecommand{\subtitle}[1]{% add subtitle to \maketitle
  \apptocmd{\@title}{\par {\large #1 \par}}{}{}
}
\makeatother
\subtitle{Toward a Comprehensive Approach to Artificial Intelligence and Labour}
\author{Julian Posada\\
\footnotesize University of Toronto}
\date{}

\begin{document}
\maketitle

\hypertarget{abstract}{%
\section{Abstract}\label{abstract}}

This commentary traces contemporary discourses on the relationship
between artificial intelligence and labour and explains why these
principles must be comprehensive in their approach to labour and AI.
First, the commentary asserts that ethical frameworks in AI alone are
not enough to guarantee workers' rights since they lack enforcement
mechanisms and the representation of different stakeholders. Secondly,
it argues that current discussions on AI and labour focus on the
deployment of these technologies in the workplace but ignore the
essential role of human labour in their development, particularly in the
different cases of outsourced labour around the world. Finally, the
commentary recommends using existing human rights frameworks for working
conditions to provide more comprehensive ethical principles and
regulations. The commentary concludes by arguing that the central
question regarding the future of work should not be whether intelligent
machines will replace humans, but who will own these systems and have a
say in their development and operation.

\hypertarget{published-as}{%
\subsubsection{Published as}\label{published-as}}

Posada, Julian. 2020. ``The Future of Work Is Here: Toward a
Comprehensive Approach to Artificial Intelligence and Labour.''
\emph{Ethics of AI in Context}

\hypertarget{introduction}{%
\section{Introduction}\label{introduction}}

\openup 0.5em In recent years, government and policy organizations,
private companies, and research institutions have explored the
implications of artificial intelligence systems in society. The
applications of AI in labour contexts have attracted particular interest
since it remains unclear how AI will impact automation and existing
working conditions. For a while, several publications that deal with the
ethics of artificial intelligence in social contexts have covered some
of the current concerns regarding this shift. These discussions are
important because of the influence these technological developments have
on society, and may inform ``our views about values and priorities, good
behaviour, and what sort of innovation is sustainable but socially
preferable'' (Floridi et al. 2018).

Canadian institutions have published frameworks that address the ethics
of artificial intelligence, such as the ``Toronto Declaration''
(Bacciareli et al. 2018) and the ``Déclaration de Montréal'' (Dilhac,
Abrassart, and Voarino 2018). At the same time, representatives from the
country have participated in the ``Artificial Intelligence in Society''
report by the OCDE (2019) and UNI's ``Top 10 Principles for Ethical
Artificial Intelligence'' (2017). The Canadian Government also released
a ``Pan-Canadian Artificial Intelligence Strategy'' through the Canadian
Institute for Advanced Research (Barron et al. 2019), one of the
twenty-eight national strategies on artificial intelligence coming from
governments in Latin America, North America, Europe, the Middle East,
and South-East Asia (Kung 2020). These documents converge in many
regards, notably the idea that AI should serve the public good,
especially around the areas of privacy, accountability, safety and
security, transparency and explainability, fairness and
non-discrimination, the human control of technology, the responsibility
of professionals, and the promotion of human values (Fjeld et al. 2020;
Millar et al. 2018).

While there is convergence concerning these broad themes, there is no
unanimity on the critical ethical AI principles. Furthermore, there is a
lack of representation from areas such as Africa and parts of Latin
America and Asia, and the interpretation of these principles, their
importance, their implementation, and the involvement of different
stakeholders generate disagreements (Jobin, Ienca, and Vayena 2019).
Ethics are fundamental to understanding the implications of
technological development and its implementation, and to what extent it
can serve workers. However, while ethics are a cornerstone in achieving
positive outcomes, their lack of enforcement mechanisms makes their
implementation challenging. Thus, governance mechanisms that regroup
different stakeholders, including non-governmental organizations,
industry, and governments, should complement ethical frameworks (Abbott
and Snidal 2009: 52). Moreover, governments should not delegate their
responsibilities to industry; the regulation of AI systems requires
effective policies (Calo 2017).

However, ethics, governance, and policy require a comprehensive view of
labour to develop a framework that addresses its connection with
artificial intelligence. Most of the declarations, reports, and
government-led strategies focus exclusively on the deployment or
applications of artificial intelligence in the workplace when
considering the relationship between these technologies and human
labour. However, none of these ethical principles addresses the forms of
labour required to develop and maintain artificial intelligence systems.
Besides, their ``techno-centric'' approach to AI regards human labour
only in quantitative terms, ignoring the quality of jobs, and seeing
artificial intelligence mostly as innovative, inevitable, and a sign of
progress, while ignoring the problems that its development and
deployment generate for workers (De Stefano 2020). This commentary will
discuss the implications of both instances in which AI affects human
labour and becomes influenced by it since a comprehensive understanding
of both is necessary for a discussion on the implications of AI for the
future of work.

\openup -0.5em

\hypertarget{current-concerns-with-ai-and-labour}{%
\section{Current Concerns with AI and
Labour}\label{current-concerns-with-ai-and-labour}}

\openup 0.5em

The OECD's ``Artificial Intelligence in Society'' report argues that
``AI is expected to complement humans in some tasks, replace them in
others and generate new types of work'' (2019). Setting aside any
uncertain future developments in technology, the contemporary landscape
of narrow AI systems suggests that, while these technologies have
incredible calculative potential and applications in a wide range of
domains, they are nonetheless incapable of judgement and far from
becoming ``artificial general intelligence'' (Smith 2019).

For this reason, the relationship between artificial intelligence and
human agency remains fundamental at all levels. As a result, human
labour remains essential in all stages of AI development and deployment;
``development'' here refers to the processes of creating, sustaining,
and maintaining AI systems, and ``deployment'' to all applications of AI
in the many instances of human labour. In many cases, such as in Amazon
warehouses (Delfanti 2019) or online gig work (Woodcock and Graham
2020), these two instances happen simultaneously, as workers support AI
development while being subject to its applications.

\hypertarget{ai-deployment}{%
\subsection{AI Deployment}\label{ai-deployment}}

The aforementioned ethical principles and strategies for AI development
are concerned about the danger of human labour being replaced by AI
systems due to the potential increase in the number of automated tasks
(CIFAR 2020; Dutton, Barron, and Boskovic 2017; Frey and Osbourne 2013;
Johal and Thirgood 2016). Benjamin Shestakofsky summarizes recent
developments in the scientific literature on contemporary automation by
distinguishing between continuity and discontinuity theories (2017).
Continuity theories predict a large-scale automation process that
threatens to replace human workers, in line with many principles and
declarations on AI. In contrast, discontinuity theories predict workers'
displacement and qualitative changes in the workplace due to the
economic expansion of AI.

Most explicit mentions of AI and labour in the ethical frameworks and
national strategies focus on mitigating the impact of automation
following discontinuity predictions. However, researchers such as David
Autor argue that, while automation threatens to replace human action
over specific tasks, it does not imply that AI will replace jobs
entirely; in fact, automation may expand the economic sectors it touches
upon and, in turn, create more jobs (2015). Antonio Casilli similarly
refers to the ``persistence of work'' despite consistent waves of
automation in history since the industrial revolution (2019: 44). In the
contemporary case, Casilli argues that the risk of particular tasks
becoming automated does not mean that the same will happen with entire
jobs (Casilli 2019), as these will become transformed, instead of
replaced.

There are several examples of how AI is actively transforming labour.
Recent studies on the impact of AI deployment in traditional workplaces
raise several concerns that are often explored in ethical discussions of
AI systems. One of the most commonly voiced concerns centres on the
discrimination fomented by hiring algorithms, since AI creates closed
systems that lack external reviewing processes, target specific
populations, and replicate the criteria of those accepted when looking
for new participants (Ajunwa and Greene 2019). As a result, individuals
from backgrounds and characteristics that do not correspond to
``optimal'' variables, notably in the case of social minorities, become
excluded by the hiring system even if they are qualified for the jobs
(Ibid.).

One of the biggest concerns regarding the deployment of artificial
intelligence in the workplace is the use of algorithmic management and
its implications for workers' agency and privacy (Adams-Prassl 2020).
Studies on the application of artificial intelligence in the workplace,
such as in productivity apps and wellness programs (Ajunwa, Crawford,
and Schultz 2017), online freelancing (Wood, Lehdonvirta, and Graham
2018), and the gig economy (Woodcock and Graham 2020), suggest high
levels of surveillance from employers. The constant need for automated
systems to collect data and quantify human behaviour promotes the
commodification of privacy (Moore and Robinson 2016), thereby allowing
it to be exchanged for employment opportunities (Ajunwa et al. 2017).
Moreover, in cases where workers rely heavily on algorithmic guidance,
the automated systems themselves become the managers, tracking and
influencing workers' actions without any accountability or transparency
(Mateescu and Nguyen 2019) and incorporating their data into improving
their systems (Delfanti 2019).

These examples suggest that the applications of artificial intelligence
in the workplace are already occurring. Therefore, the issue at hand
regarding the deployment of AI is not the reskilling of workers whose
jobs will be replaced, but the degradation of their working conditions
due to the deskilling of the labour process, the implementation of
algorithmic management, and privacy issues. Still, the situations
previously described illustrate only some of the current concerns with
AI and labour, since the development of these systems relies heavily on
human labour to exist and subsist.

\hypertarget{ai-development}{%
\subsection{AI Development}\label{ai-development}}

Researchers Kate Crawford and Vladan Joler from the AI Now Institute
published an analysis of the labour and natural resources required to
power Amazon's virtual assistant technology Alexa (2018). The authors
stress the importance of natural resources and human action to sustain
automated systems from a material perspective by arguing that, in the
context of AI, ``each small moment of convenience -- be it answering a
question, turning on a light, or playing a song -- requires a vast
planetary network, fueled by the extraction of non-renewable materials,
labour, and data'' (Crawford and Joler 2018).

The authors trace the creation---and disposal---of the material and
digital components that power the Alexa with an analysis based on the
dialectic of subject and object in the economy. They illustrate the
evolution of the device that supports the AI, from the extraction and
transformation of natural resources into electronic components to their
assembly and shipping and, then, their collection, recycling, and
disposal. Regarding the transformation of data, the authors illustrate
its constant flows, starting with the quantification and capture of
natural processes through these electronic devices, and their passage
through various infrastructures (domestic, local, national, and global).
Then, they describe how Amazon processes this data. At this level, the
authors acknowledge both the labour of engineers and technicians in
maintaining these infrastructures and the unrecognized, low-paid, or
even unpaid labour of ``digital pieceworkers'' (Dubal 2020) that
supervise, correct, and even impersonate the AI system when required
(Tubaro, Casilli, and Coville 2020) through ``crowdsourcing.''

The ``Anatomy of an AI System'' illustrates how the development of
artificial intelligence depends heavily on different types of human
labour and natural resources that span the entire planet. For the
authors, the artificial intelligence system ``becomes a complex
structure of supply chains within supply chains, a zooming fractal of
tens of thousands of suppliers, millions of kilometres of shipped
materials and hundreds of thousands of workers included within the
process even before the product is assembled on the line'' (Crawford and
Joler 2018). Thus, by focusing so heavily on the deployment of AI
systems and the ``future of work,'' the current relationship between
artificial intelligence and labour remains ignored in many cases.

Platforms have become central to the creation of AI because they allow a
permissive quantification of the natural world; there are
``(re-)programmable digital infrastructures that facilitate and shape
personalised interactions among end-users and complementors, organised
through the systematic collection, algorithmic processing, monetisation,
and circulation of data'' (Poell, Nieborg, and Dijck 2019). The
platformization process involves the ``penetration of the
infrastructures, economic processes, and government frameworks of
platforms in different economic sectors and spheres of life'' (Poell et
al. 2019), which allows an economically efficient way to harness the
data necessary for AI systems while reducing production and development
costs to a minimum. Thus, the platform model constitutes the primary
organizational paradigm for the major corporations that develop
artificial intelligence today (Casilli and Posada 2019).

In this context, labour platforms are an excellent example of the
relationship between the development and the deployment of artificial
intelligence. Despite representing a small portion of the current global
workforce (O'Farrell and Montagnier 2020), their study is fundamental to
comprehend the ethical implications of AI and work. The platformization
of labour allows firms to reduce production costs by externalizing---and
outsourcing---jobs outside of their scope to ``independent contractors''
(Prassl 2018: 79). For instance, in the case of ``digital piecework''
platforms, where unrecognized and invisibilized workers provide data and
serve as supervisors for machine learning technologies (Dubal 2020), the
platforms allow algorithms to serve as managers (Mateescu and Nguyen
2019), and to surveil and ``datafy'' (or harness as data) the behaviour
of workers (Casilli and Posada 2019).

Furthermore, as platforms serve as intermediaries, they also prevent
workers from engaging in collective action (Wood et al. 2018), even
deliberately (Posada and Shade 2020). By looking at the heavily
deregulated status of platform labour and comparing it to the ethical
principles for AI previously mentioned, it is evident how the issues of
privacy, accountability, explainability, and fairness are still not met
in these instances where the deployment and development of AI merge.

\openup -0.5em

\hypertarget{human-rights-based-regulations}{%
\section{Human Rights-Based
Regulations}\label{human-rights-based-regulations}}

\openup 0.5em

Echoing the debate between continuity and discontinuity theories, Ken
Goldberg argues that intelligent machines will work closely with humans
instead of replacing them, a concept that he calls ``multiplicity''
(Bauer 2018). As the total quantification of human and social experience
remains a long term---or even impossible---dream, the labour of those
required to develop AI---and those affected by its application---will
remain central to the discussion of its ethics, governance, and policy.
In the context of ``multiplicity,'' the central issue in terms of labour
relations will remain the ownership, fairness, and power relations
between those who control these automated systems and those considered
its ``users.''

Emphasizing the limits of ethical principles, Yeung, Howes, and Pogrebna
warn against their lack of enforcement and the immense influence of
corporate entities in their development (2020). The authors suggest
using international human rights standards as a basis for AI's ethical
frameworks, as they are ``grounded on a shared commitment to uphold the
inherent human dignity of each and every person'' (Yeung et al. 2020).
In a similar vein, Valerio de Stefano argues that a human-rights
approach to the regulation of AI labour limits and rationalizes the
``exercise of managerial prerogatives'' that can affect the autonomy and
dignity of workers (De Stefano 2020: 16).

There are several historical human rights conventions related to labour
that address workers' concerns in the development and application of
artificial intelligence better than the recently published principles
and strategies do. Through the International Labour Organization, the
United Nations has issued several conventions on human rights related to
labour issues. These fundamental human labour rights include the freedom
of association and recognition of the right to collective bargaining,
the elimination of forced labour, the abolition of child labour and
discrimination, and equal remuneration (ILO 1998). As mentioned, there
is still a long way to go when it comes to upholding these basic labour
principles in the development and deployment of AI. For example, some
platforms deliberately hinder collective bargaining by their workers
(Woodcock and Graham 2020) and the application of artificial
intelligence in worker recruitment reproduces biases towards particular
social minorities (Ajunwa et al. 2017).

Furthermore, the proposed idea of ``regulatory markets'' can complement
existing regulations, notably at an international level. Current
government regulations on artificial intelligence are national in scope
and fail to address the rapid development and deployment of these
systems (Clark and Hadfield 2019). This situation becomes problematic
when online labour markets include several countries and create a
situation where workers, intermediaries, and companies are subject to
several jurisdictions (Graham, Hjorth, and Lehdonvirta 2017). Clark and
Hadfield propose a situation where international ``private regulators''
(which could be called ``independent regulators'' instead) serve
national governments by overseeing compliance with governmental
principles and desired outcomes (2019).

Currently, the ``FairWork Foundation'' operates similarly to these
proposed regulators in partnership with the International Labour
Organization. Placed between consumers and companies and modelled after
the Fair-Trade Movement, the FairWork Foundation evaluates digital
labour platforms according to the principles of fair pay, conditions,
contracts, management, and representation (fair.work). The scores inform
workers, clients, and the public-at-large about the quality of work in
these platforms, hoping that ``core transparent production networks can
lead to better working conditions for digital workers around the world''
(Graham and Woodcock 2018: 250-251). Like Clark and Hadfield's
regulators, this initiative complements existing ethical principles,
direct government regulations, and independent collective action and
organization, which remain fundamental in ensuring that the development
and deployment of artificial intelligence serve the public good.

\openup -0.5em

\hypertarget{conclusion}{%
\section{Conclusion}\label{conclusion}}

\openup 0.5em

Ethical principles are essential to the relationship between AI and
human labour, but they need to become more comprehensive. Principles
alone cannot be enforced in social contexts, and they should go hand in
hand with clear governance procedures involving multiple parties and
oversight from international and national policies that aim to respect
already established human rights. These actions cannot focus solely on
the prospective ``future of work'' or on the deployment of artificial
intelligence in the workplace. AI is here, and it is already impacting
the ``present of work.'' Machines depend on humans to exist, and both
entities are already complementing each other. Ethics, governance,
policy, collective action, and other alternatives will remain essential,
as the question will not be if machines will be replacing humans, but
who will own the machines and have a say in the relationship between
them and humans.

\openup -0.5em

\hypertarget{references}{%
\section*{References}\label{references}}
\addcontentsline{toc}{section}{References}

\hypertarget{refs}{}
\leavevmode\hypertarget{ref-Abbott2009}{}%
Abbott, Kenneth W. and Duncan Snidal. 2009. ``The Governance Triangle:
Regulatory Standards Institutions and the Shadow of the State.'' Pp.
44--88 in \emph{The politics of global regulation}, edited by W. Mattli
and N. Woods. Princeton, NJ: Princeton University Press.

\leavevmode\hypertarget{ref-Adams-Prassl2020}{}%
Adams-Prassl, Jeremias. 2020. ``When your Boss comes Home: Three Fault
Lines for the Future of Work in the Age of Automation, AI, and
COVID-19.'' \emph{Ethics of AI in Context} 1--11.

\leavevmode\hypertarget{ref-Ajunwa2017}{}%
Ajunwa, Ifeoma, Kate Crawford, and Jason M. Schultz. 2017. ``Limitless
worker surveillance.'' \emph{California Law Review} 105(3):735--76.

\leavevmode\hypertarget{ref-Ajunwa2019a}{}%
Ajunwa, Ifeoma and Daniel Greene. 2019. ``Platforms at work: Automated
hiring platforms and other new intermediaries in the organization of
work.'' \emph{Research in the Sociology of Work} 33:61--91.

\leavevmode\hypertarget{ref-Autor2015}{}%
Autor, David H. 2015. ``Why Are There Still So Many Jobs? The History
and Future of Workplace Automation.'' \emph{Journal of Economic
Perspectives} 29(3):3--30.

\leavevmode\hypertarget{ref-Bacciareli2018}{}%
Bacciareli, Anna, Joe Westby, Estelle Massé, Drew Mitnick, Fanny
Hidvegi, Boye Adegoke, Frederike Kaltheuner, Malavika Jayaram, Yasodara
Córdova, Solon Barocas, and William Isaac. 2018. \emph{The Toronto
Declaration : Protecting the rights to equality and non-discrimination
in machine learning systems}. Toronto, ON: Amnesty International; Access
Now.

\leavevmode\hypertarget{ref-Barron2019}{}%
Barron, Brent, Nabilah Chowdhury, Krista Davidson, and Kurt Kleiner.
2019. \emph{Annual Report of the CIFAR Pan-Canadian AI Strategy}. edited
by E. Strome. Ottawa, ON: Canadian Institute for Advanced Research
CIFAR.

\leavevmode\hypertarget{ref-Bauer2018}{}%
Bauer, Lisa. 2018. ``Multiplicity Not Singularity: Ken Goldberg on the
Future of Work.'' \emph{Blum Center for Developing Economies}.

\leavevmode\hypertarget{ref-Calo2017}{}%
Calo, Ryan. 2017. ``Artificial Intelligence Policy: A Roadmap.''
\emph{SSRN Electronic Journal} 1--28.

\leavevmode\hypertarget{ref-Casilli2019a}{}%
Casilli, Antonio A. 2019. \emph{En attendant les robots}. Paris:
Éditions du Seuil.

\leavevmode\hypertarget{ref-Casilli2019}{}%
Casilli, Antonio A. and Julian Posada. 2019. ``The Platformisation of
Labor and Society.'' in \emph{Society and the internet}, edited by M.
Graham and W. H. Dutton. Oxford: Oxford University Press.

\leavevmode\hypertarget{ref-CIFAR2020}{}%
CIFAR. 2020. \emph{Report on Canada-U.S. AI Symposium on Economic
Innovation}. Toronto, ON: Canadian Institute for Advanced Research
CIFAR.

\leavevmode\hypertarget{ref-Clark2019}{}%
Clark, Jack and Gillian K. Hadfield. 2019. ``Regulatory Markets for AI
Safety.'' \emph{arXiv} 1--23.

\leavevmode\hypertarget{ref-Crawford2018}{}%
Crawford, Kate and Vladan Joler. 2018. \emph{Anatomy of an AI system}.
New York, NY: AI Now Institute.

\leavevmode\hypertarget{ref-Delfanti2019}{}%
Delfanti, Alessandro. 2019. ``Machinic dispossession and augmented
despotism: Digital work in an Amazon warehouse.'' \emph{New Media \&
Society}.

\leavevmode\hypertarget{ref-DeStefano2020}{}%
De Stefano, Valerio. 2020. ``Algorithmic Bosses and How to Tame Them.''
\emph{Ethics of AI in Context}.

\leavevmode\hypertarget{ref-Dilhac2018}{}%
Dilhac, Marc-Antoine, Christophe Abrassart, and Nathalie Voarino. 2018.
\emph{Rapport de la déclaration de Montréal pour un développement
responsable de l'intelligence artificielle}. Montréal, QC: Déclaration
de Montréal IA responsable.

\leavevmode\hypertarget{ref-Dubal2020}{}%
Dubal, Veena B. 2020. ``The Time Politics of Home-Based Digital
Piecework.'' \emph{Ethics of AI in Context}.

\leavevmode\hypertarget{ref-Dutton2017}{}%
Dutton, Tim, Brent Barron, and Gaga Boskovic. 2017. \emph{Building an AI
world - Report on National and Regional AI Strategies}. Toronto, ON:
CIFAR.

\leavevmode\hypertarget{ref-Fjeld2020}{}%
Fjeld, Jessica, Nele Achten, Hannah Hilligoss, Adam Christopher Nagi,
and Madhulika Srikumar. 2020. \emph{Principled Artificial Intelligence}.
Cambridge, MA: Berkman Klein Center for Internet \& Society.

\leavevmode\hypertarget{ref-Floridi2018}{}%
Floridi, Luciano, Josh Cowls, Monica Beltrametti, Raja Chatila, Patrice
Chazerand, Virginia Dignum, Christoph Luetge, Robert Madelin, Ugo
Pagallo, Francesca Rossi, Burkhard Schafer, Peggy Valcke, and Effy
Vayena. 2018. ``AI4People---An Ethical Framework for a Good AI Society:
Opportunities, Risks, Principles, and Recommendations.'' \emph{Minds and
Machines} 28(4):689--707.

\leavevmode\hypertarget{ref-Frey2013}{}%
Frey, Carl Benedikt and Michael A. Osbourne. 2013. \emph{The Future of
Employment: How Susceptible Are Jobs to Computerisation?} Oxford: The
Oxford Martin Program on Technology; Employment.

\leavevmode\hypertarget{ref-Graham2017}{}%
Graham, Mark, Isis Hjorth, and Vili Lehdonvirta. 2017. ``Digital labour
and development: impacts of global digital labour platforms and the gig
economy on worker livelihoods.'' \emph{Transfer: European Review of
Labour and Research} 23(X):1--28.

\leavevmode\hypertarget{ref-Graham2018b}{}%
Graham, Mark and Jamie Woodcock. 2018. ``Towards a Fairer Platform
Economy: Introducing the Fairwork Foundation.'' \emph{Alternate Routes}
29:242--53.

\leavevmode\hypertarget{ref-ILO1998}{}%
ILO. 1998. \emph{ILO Declaration on Fundamental Principles and Rights
and Rights at Work and Its Follow Up}. Geneva: International Labour
Organization.

\leavevmode\hypertarget{ref-Jobin2019}{}%
Jobin, Anna, Marcello Ienca, and Effy Vayena. 2019. ``The global
landscape of AI ethics guidelines.'' \emph{Nature Machine Intelligence}
1(9):389--99.

\leavevmode\hypertarget{ref-Johal2016}{}%
Johal, Sunil and Jordann Thirgood. 2016. \emph{Working Without a Net.
Rethinking Canada's social policy in the new age of work}. 132. Toronto,
ON: Mowat Centre.

\leavevmode\hypertarget{ref-Kung2020}{}%
Kung, Johnny. 2020. \emph{Building an AI world: Report on National and
Regional AI Strategies}. Toronto, ON: Canadian Institute for Advanced
Research CIFAR.

\leavevmode\hypertarget{ref-Mateescu2019}{}%
Mateescu, Alexandra and Aiha Nguyen. 2019. \emph{Algorithmic Management
in the Workplace}. New York, NY: Data \& Society Research Institute.

\leavevmode\hypertarget{ref-Millar2018}{}%
Millar, Jason, Brent Barron, Koichi Hori, Rebecca Finlay, Kentaro
Kotsuki, and Ian Kerr. 2018. \emph{Accountability in AI. Promoting
Greater Social Trust Acknowledgements and Authors Note}. Toronto, ON:
Canadian Institute for Advanced Research CIFAR; Institute for
Information; Commnucations Policy.

\leavevmode\hypertarget{ref-Moore2016}{}%
Moore, Phoebe and Andrew Robinson. 2016. ``The quantified self: What
counts in the neoliberal workplace.'' \emph{New Media \& Society}
18(11):2774--92.

\leavevmode\hypertarget{ref-OECD2019a}{}%
OECD. 2019. \emph{Artificial Intelligence in Society}. Paris: OECD
Publishing.

\leavevmode\hypertarget{ref-OFarrell2019}{}%
O'Farrell, Rory and Pierre Montagnier. 2020. ``Measuring digital
platform‐mediated workers.'' \emph{New Technology, Work and Employment}
35(1):130--44.

\leavevmode\hypertarget{ref-Poell2019}{}%
Poell, Thomas, David B. Nieborg, and José van Dijck. 2019.
``Platformisation.'' \emph{Internet Policy Review} 8(4).

\leavevmode\hypertarget{ref-Posada2020}{}%
Posada, Julian and Leslie Regan Shade. 2020. ``Platform Labour
Discourse: How Hyr Targets the `Bucket List Generation'.''
\emph{Democratic Communiqué} 29(1):78--96.

\leavevmode\hypertarget{ref-Prassl2018a}{}%
Prassl, Jeremias. 2018. \emph{Humans as a Service: The Promise and
Perils of Work in the Gig Economy}. Oxford: Oxford University Press.

\leavevmode\hypertarget{ref-Shestakofsky2017a}{}%
Shestakofsky, Benjamin. 2017. ``Working Algorithms: Software Automation
and the Future of Work.'' \emph{Work and Occupations} 44(4):376--423.

\leavevmode\hypertarget{ref-Smith2019}{}%
Smith, Brian Cantwell. 2019. \emph{The Promise of Artificial
Intelligence: Reckoning and Judgment}. Cambridge, MA: MIT Press.

\leavevmode\hypertarget{ref-Tubaro2020}{}%
Tubaro, Paola, Antonio A. Casilli, and Marion Coville. 2020. ``The
trainer, the verifier, the imitator: Three ways in which human platform
workers support artificial intelligence.'' \emph{Big Data \& Society}
7(1).

\leavevmode\hypertarget{ref-UNIGlobalUnion2017}{}%
UNI Global Union. 2017. \emph{Top 10 Principles for Ethical Artificial
Intelligence}. Lyon: UNI Global Union.

\leavevmode\hypertarget{ref-Wood2018d}{}%
Wood, Alex J., Vili Lehdonvirta, and Mark Graham. 2018. ``Workers of the
Internet unite? Online freelancer organisation among remote gig economy
workers in six Asian and African countries.'' \emph{New Technology, Work
and Employment} 33(2):95--112.

\leavevmode\hypertarget{ref-Woodcock2020}{}%
Woodcock, Jamie and Mark Graham. 2020. \emph{The Gig Economy: A Critical
Introduction}. London: Polity Press.

\leavevmode\hypertarget{ref-Yeung2020}{}%
Yeung, Karen, Andrew Howes, and Ganna Pogrebna. 2020. ``AI Governance by
Human Rights-Centred Design, Deliberation and Oversight: An End to
Ethics Washing.'' in \emph{The oxford handbook of ai ethics}. Oxford:
Oxford University Press.

\end{document}